\documentclass{aa}

\usepackage{graphicx}

\usepackage{txfonts}

\usepackage{natbib}
\usepackage{color}
\usepackage{fancyvrb}

\title{Using birefringent elements and imaging Michelsons for calibration of high precision planet finding spectrographs}

\author{J. Schou}

\date{\today}

\institute{
Max-Planck-Institut f\"ur Sonnensystemforschung, Justus-von-Liebig-Weg
3, 37077 G\"ottingen, Germany\\
\email{schou@mps.mpg.de}
}

\begin{document} 

 
\abstract
{
One of the main methods used for finding extrasolar planets is the radial
velocity technique, in which the Doppler shift of a star due to an orbiting planet is measured.
These measurements are typically performed using cross-dispersed echelle spectrographs. Unfortunately such spectrographs are large and expensive and their accurate calibration continues to be challenging.
}
{
The aim is to develop a different way to provide a calibration signal.
}
{
A commonly used way to introduce a calibration signal is to insert an iodine cell in the beam. Disadvantages of this include that the lines are narrow, do not cover the entire spectrum and that light is absorbed.
Here I show that inserting a birefringent element or an imaging Michelson, combined with Wollaston prisms eliminates these three shortcomings, while maintaining most of the benefits of the iodine approach.
}
{
The proposed designs can be made very compact, thereby providing a convenient way of calibrating a spectrograph.
Similar to the iodine cell approach, the calibration signal travels with the stellar signal, thereby reducing the sensitivity to spectrograph stability.
The imposed signal covers the entire visible range and any temperature drifts will be consistent and describable by a single number.
Based on experience with similar devices used, in a different configuration, by the Helioseismic and Magnetic Imager, it is shown that the calibration device can be made stable at the 0.1~m/s level, over a significant wavelength range, on short to medium time scales.
}
{
While promising, many details still need to be worked out. In particular a number of laboratory measurements are required in order to finalize a design and estimate actual performance and it would be desirable to make a proof of concept.
}


\maketitle


\section{Introduction}
The design of spectrographs for highly precise and accurate Doppler measurements
for planet finding and asteroseismology is driven by several different
scientific requirements, the simultaneous satisfaction of which has resulted
in very large and expensive instruments.
Driving requirements include the efficient use of the photons collected
by the telescope, a desire for extreme stability, and a need to match
available detectors.
These requirements have led to almost all such spectrographs being some
variation of a cross-dispersed echelle design \citep{2015sifg.book.....E}.
Indeed, 51 Pegasi, the first planet discovered by the radial velocity method, was originally detected using such spectrographs:
ELODIE \citep{1996A&AS..119..373B}, Advanced Fiber Optic Echelle (AFOE) \citep{1994PASP..106.1285B} and the Hamilton Echelle Spectrometer \citep{1987PASP...99.1214V}.

A persistent problem with such spectrographs is that their inherent stability is not adequate for detecting Earth sized planets in the habitable zone
around Sun-like stars, which typically cause a stellar reflex motion of $\approx$0.1~m/s.
A review of the challenges involved in this have been discussed by \cite{2016PASP..128f6001F}, which discusses various methods currently in use.
These methods generally involve
introducing a calibration signal into the spectrograph.
One approach is to introduce absorption lines typically using an iodine cell \citep{1990ASPC....8..335B, 1996PASP..108..500B}.
Another approach is to use emission lines,
which are typically generated using a ThAr lamp \citep{1990ASPC....8..335B,2007A&A...468.1115L}.
Alternatively a line spectrum may be generated using a laser comb \citep{2012OExpr..2013711P,2012Natur.485..611W,2019Optic...6..233M}.
Finally a reference spectrum may be generated using a Fabry-P\'erot etalon \citep{2010SPIE.7735E..4XW,2014A&A...569A..77R,2015A&A...581A.117B,2015PASP..127..880S,Jennings:20}.
In the case of iodine cells and thorium-argon (ThAr) lamps, the wavelengths of the spectral lines are inherently stable. Laser combs and Fabry-P\'erots are referenced to an external reference.

The calibration signal can either be superimposed onto the stellar light or introduced in a separate beam, effectively creating additional orders on the detector.
The former approach has the disadvantage of removing signal photons or adding photons without signal, in both cases degrading the data.
The latter approach has the disadvantage that the stellar and calibration signals do not travel together and thus do not have the same sensitivity to instrumental drifts.
Also, a larger CCD is required than for the former approach, due to the doubled orders.

Additional problems include that while the wavelengths are very stable the line strengths are not, leading to problems if the lines are unresolved; that it is difficult to cover the entire wavelength range; and that some of the lines from e.g. ThAr lamps are much brighter than others, leading to saturation and bleeding into different orders.

Desirable features of a calibration signal thus include not degrading the S/N (no photons added or subtracted), covering the entire spectrum, that the stellar light and calibration signal follow each other through the relevant parts of the instrument, and that it is itself inherently stable or can be easily calibrated.
None of the designs mentioned above have all of these features.

For solar observations the requirements are quite different.
Here there is a need to image an extended object with good spatial
resolution in a variety of variables, including Doppler shift.
On the other hand, photons are abundant and as a consequence
it is possible to use a narrow wavelength range and to accept substantial
photon losses.
For making Doppler images, this has typically led to the use of so-called
"filtergraphs". In these, images are taken through narrow-band tunable filters, which are realized
by various combinations of dielectric filters, birefringent (Lyot) elements,
Michelson interferometers, and Fabry-P\'erot elements.
Examples include the ground based Global Network Oscillations Group \citep[GONG;][]{1996Sci...272.1284H},
the Michelson Doppler Imager \citep[MDI;][]{1995SoPh..162..129S} onboard
the Solar and Heliospheric Observatory \citep{1995SoPh..162....1D},
the Helioseismic and Magnetic Imager \citep[HMI;][]{2012SoPh..275..229S}
onboard the Solar Dynamics Observatory \citep{2012SoPh..275....3P}
and the Polarimetric and Helioseismic Imager \citep{2020A&A...642A..11S} onboard Solar Orbiter \citep{2021A&A...646A.121G}, each of which use a different combination of filters.
For making spatially resolved Doppler images of giant planets, Mach-Zehnder interferometers, which are somewhat similar to the Michelsons mentioned above, have been used \citep{2007A&A...474.1073S,2019Icar..319..795G}.

In Sect. \ref{sec:filters} I briefly discuss the operation of
birefringent filters and wide field imaging Michelsons.
In Sect. \ref{sec:simple} I show how such a device can be
incorporated into a spectrograph, how this may be used
to make a compact design, and provide a good calibration.
In Sect. \ref{sec:complicated} a generalization of the
design is explored and it is shown how this can be used to improve the performance and/or decrease the size of a spectrograph.

Sect. \ref{sec:conclusion} provides a summary.
The present paper is only intended to show a concept,
thus a detailed layout of a spectrograph is not presented. Nonetheless, a number of practical details are discussed in Appendix \ref{practical}.

It should be noted that some of the ideas discussed here have been
discussed in connection with 
the Externally Dispersed Interferometer
\cite[EDI;][]{2014SPIE.9147E..17E, 2012SPIE.8446E..4JE, 2011SPIE.8146E..0ME} and 
the Dispersed Fixed-Delay Interferometer \cite[DFDI;][]{2010ApJS..189..156V}
which incorporate a Michelson interferometer.
For a detailed theoretical understanding of these, see \cite{2010ApJS..189..156V}, who also provide an extensive list of references.

\section{Basics of imaging filters}
\label{sec:filters}
In the solar case imaging filters, that is filters through which an extended object can be imaged in a narrow passband, are used. 
Many different types of imaging filters 
have been used,
including dielectric interference filters, Fabry-P\'erot filters,
birefringent filters, and Michelson interferometers.
In the following subsections I briefly discuss the latter two, as they are particularly
relevant for the present purpose.
For details of these types of filters, their use in HMI, and references to more detailed papers, see \cite{Couvidat2012}.

\subsection{Birefringent filters}
\label{sec:biref}
A simple birefringent filter element consists of
a piece of birefringent material followed by an exit polarizer.
The input light must be polarized, typically either circularly or linearly at 45$^\circ$
to the fast and slow axis of the birefringent element.
Various birefringent materials can be used, with calcite being
the most common in Lyot filters for solar use, often combined with
KDP (KH$_2$PO$_4$) or ADP (NH$_4$H$_2$PO$_4$) for temperature compensation.
Recently LiNbO$_3$ has also been studied by 
\cite{2016JGRA..121.6184T}, who also list other candidate materials.
Of these $\alpha-$BaB$_2$O$_4$, TeO$_2$, TiO$_2$ and YVO$_4$ appear to be good candidates, but have not been used in this context.
However, before these materials are used their
optical quality and the temperature and wavelength dependence of their birefringence will need to be determined.

Locally (i.e. ignoring secular variations), the transmission
as a function of wavelength is given by
\begin{equation}
\label{eq:trans0}
T(\lambda)=\frac{1+{\rm B}\cos(2\pi(\lambda-\lambda_0)/{\rm FSR})}{2},
\end{equation}
where $\lambda$ is the wavelength, B is the contrast, FSR is the so-called free spectral range
(determined by the length and birefringence of the element),
and $\lambda_0$ is the position of a transmission peak.
If the entrance or exit polarization is changed
by 90$^\circ$ (or between left circular polarization and right circular polarization), $\lambda_0$ is changed by FSR/2, that is the peaks and troughs
of Eq. (\ref{eq:trans0}) are exchanged.
Another factor of 0.5 needs to multiplied on the transmission for
the first element in a series in order to take into account the
loss of light in the input polarizer.
In reality, B<100\% and the FSR will change slowly with wavelength. However, as discussed in Secs. \ref{sec:noise} and \ref{sec:calnoise}, these issues are of little consequence.  

An important property of a birefringent element is that all the
pieces can be greased (oiled) together, resulting in low reflective
losses and a lack of sensitivity to air pressure.
One may also choose to temperature compensate an element by using
two different materials (in the case of HMI calcite+ADP
or calcite+KDP depending on the element, as described in Sec. \ref{sec:temp_stab}).
For a simple element the central wavelength depends significantly on both the angle to normal and the azimuthal direction relative to the optical axis.
For solar imaging applications the elements are thus typically wide-fielded by using two
crossed half elements with a 1/2~wave plate in between, which results in a lower dependence on the incidence angle and removes the azimuthal dependence.

A Lyot filter is constructed by placing several birefringent elements,
generally with a factor of two in the FSR, in series, using the
fact that the exit polarizer of one element can act as the
entrance polarizer of the next element and tuning each to the
design wavelength.
Using such a design with ideal polarizers the transmission is 0.5
at the design wavelength, regardless of the number of elements.

While a Lyot filter is typically optimized to work at a single wavelength, a single element will work over a very large wavelength range, only limited by the transmission of the material used, which in the case of calcite ranges from 4000\AA\ to about $2 \mu m$.
If all the elements in a Lyot filter are made tunable, resulting in what is known as a Universal Birefringent Filter, the entire transmission range can, in principle, be covered.
To accomplish the tuning a 1/4~wave plate and a rotatable
1/2~wave plate are added to each element between one of the polarizers and the birefringent
crystal. This allows one to tune $\lambda_0$ of
the element and obtain a transmission peak at any desired wavelength.

HMI uses a Lyot with five elements (E1 through E5) with nominal FSRs of 690~m\AA, 1380~m\AA, 2758~m\AA, 5516~m\AA, and 11\,032~m\AA. Of these E1 is tunable.
The average contrasts are all $\ge$ 96\%.

\subsection{Imaging Michelsons}
\label{sec:michelsons}
An alternative type of imaging filters is polarizing Michelsons \citep{1980ApOpt..19.2046T,2010ISSIR...9..327T},
such as those used by GONG, MDI, and HMI.
In these the input light is divided into two
polarizations in a polarizing beamsplitter, reflected in the two arms
and recombined in the polarizing beamsplitter.
With the appropriate addition of various waveplates, the light
exits the fourth face of the beamsplitter.

While very different in design, the properties are quite similar
to those of the birefringent filters described in the previous section.
In particular the transmission profile is also given by Eq. (\ref{eq:trans0}),
they can be tuned by rotating a half wave plate and be temperature
compensated.
Main differences include that the light is reflected by 90$^\circ$, a possibility of an improved angular dependence of the central wavelength (fourth order instead of second order), that they
can often be made smaller, and do not depend on difficult to obtain materials.
Downsides include that, at least for the design used for MDI and HMI,
one of the legs is air/vacuum spaced meaning that they are
highly sensitive to air pressure (if not operated in vacuum) and temperature.
Temperature compensation using solid legs with, e.g., two different types of glass, has not been used and would require careful attention to mechanical stresses.

HMI contains two Michelsons, the so-called Wide Band (WB) Michelson with an FSR of 344~m\AA\ and the Narrow Band (NB) Michelson with an FSR of 172~m\AA .
Both of these are tunable. As for the Lyot element this is done using a rotating half wave plate.
In a stellar context, imaging and temperature compensated, but non-polarizing, Michelsons have also been investigated in the context of the EDI and DFDI designs \citep{2008PASP..120.1001M,2010SPIE.7734E..3NW,2012PASP..124..598W}.

\section{A simple design}
\label{sec:simple}
The idea of an EDI is basically to insert a classical Michelson 
into the collimated beam of a standard spectrograph, in order to increase the effective resolution of the spectrograph.
A downside of such an arrangement is that half of the light is lost.

\begin{figure}
\begin{center}
\includegraphics[width=0.95\columnwidth]{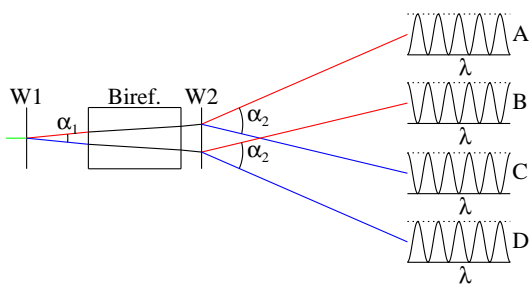}
\end{center}
\caption[]{
A sketch of the simple design. Unpolarized (green) light enters on
the left. This is split into vertically (red) and horizontal (blue)
polarized light by  the first Wollaston prism (W1). 
Inside the birefringent element the polarization of the light depends
on wavelength and position.
After the birefringent element the light is split by the second
Wollaston prism (W2), with twice the offset angle of the first.
The transmissions of the four beams, as a function of wavelength,
are shown on the right.
Note that the modulation of the top (A) and bottom (D) beams are identical,
as are the two middle ones (B and C).
For illustrative purposes the relative distances are not to scale,
the angles have been vastly exaggerated and are not two-to-one, and only the
center ray of each collimated beam has been shown.
In reality the collimated beam would nearly fill the element, the spacing of the rays would be of the order of the fiber diameter and $\alpha_2 \approx 2 \alpha_1$ or $\alpha_2 \approx \alpha_1/2$.
For clarity the Wollaston prisms have been spaced away from the element. In reality they would likely be in contact. Also, they have been shown with zero thickness.
}
\label{fig:one-element}
\end{figure}

\begin{figure}
\begin{center}
\includegraphics[width=0.95\columnwidth]{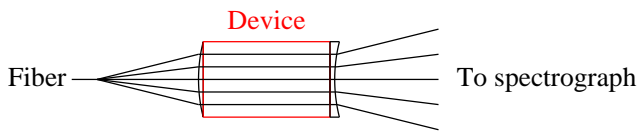}
\end{center}
\caption[]{
One possible way to insert the device into a spectrograph.
Here it is inserted in a short section of collimated beam close to the fiber.
Alternatively, the light can be refocused after the device.
}
\label{fig:overall}
\end{figure}

Clearly it would desirable to make use of all the light.
The solution proposed here is to split the input light into two complementary polarization states and to replace the classical Michelson with one of the imaging filters described in the previous section.
A sketch of the proposed design is shown in Fig. \ref{fig:one-element}.
There are multiple options for how to incorporate the device into a spectrograph, one of which is shown in Fig. \ref{fig:overall}. In that setup the device is placed in a short section of collimated light immediately after the fiber. After this the light can either be focused back to a point or continue as it would have without the device. The latter option is shown here.
Given that they are likely to be the preferred option and given that they make for simpler drawings, I have chosen to use a birefringent element for the illustrations rather than a Michelson interferometer.

The splitting of the light into horizontal and vertical polarization can be achieved using a so-called Wollaston prism. Such a prism consists of two oppositely oriented prisms of a birefringent material with crossed optical axes cemented together. When light is passed through such a prism, the horizontally polarized part of the light will be refracted up by some angle and the vertically polarized part down by the same angle. By choosing the angle of the prisms appropriately the divergence angle can be set to the desired value.

As discussed in the previous section, the transmission of the two beams, at a fixed output polarization, will be oppositely modulated. One will be modulated by
the transmission in Eq. (\ref{eq:trans0}) with one
value of $\lambda_0$ and one with FSR/2 added to $\lambda_0$ (in other words with peaks and troughs exchanged).

Clearly, using a polarizer on the output will result in the loss of half the light. 
To get around this problem the output polarizer can also be replaced by a Wollaston prism.
Again, this will cause each of the beams inside the element to be split in two beams with opposite polarization and opposite modulation. In total there will be four output beams, two with the one modulation pattern and two with the opposite, as illustrated in Fig. \ref{fig:one-element}.
Examples of the resulting spectra are shown in Fig. \ref{fig:spec1}.
Basically each of these spectra is similar to a spectrum taken through an iodine cell, with the irregular iodine spectrum replaced by a regular sinusoidal modulation.
The spectra are even more similar to those obtained by inserting a Fabry-P\'erot into the beam \citep{2010SPIE.7735E..4XW,2015A&A...581A.117B}.
However, unlike in those approaches, no light is lost. The sum of the four spectra contains all the input light.

In order for the four beams to remain separate, the deflection angle of the second Wollaston prism 
($\alpha_2$) must be chosen to be different from that of the first prism ($\alpha_1$), most simply by setting $\alpha_2 \approx 2 \alpha_1$ or $\alpha_2 \approx \alpha_1/2$.
The smaller the two angles must be chosen such that the spectra are spaced by at least their widths, that is by at least a fiber width.
Since the distance from the fiber exit to the device would be very roughly similar to the length of the device, it follows that the offset of the beams at the exit will very roughly be one fiber diameter.
Regarding the choice of $\alpha_2 \approx 2 \alpha_1$ versus $\alpha_2 \approx \alpha_1/2$, this will depend on a more detailed analysis. The former has the advantage of minimizing the angles in the device, thereby minimizing any angular effects and thermal gradients. The latter has the advantage of placing the complementary spectra (A,C and B,D) next to each other.

One way to look at the setup in Fig. \ref{fig:overall} is to note that if one were to look into the device, one would see the  fiber replicated four times above each other, each representing one of the output beams.
This is similar to the well-known double images seen through calcite crystals.

This design has several advantages over a standard
cross dispersed echelle design, some of which have been pointed
out in connection with the EDI and DFDI designs:
\begin{itemize}
\item The modulation provides a convenient and consistent calibration signal
across the spectrum.
\item The modulation signal can be made extremely stable and is
easy to calibrate.
\item It increases the efficiency of the spectrograph in the sense that
a lower noise level on the Doppler shift can be obtained with a
spectrograph of the same resolution. Alternatively the same
noise performance can be achieved with a lower spectrograph
resolution.
\end{itemize}
Below I will address these and
discuss some advantages and disadvantages.

\subsection{Noise performance}
\label{sec:noise}

\begin{figure}
\begin{center}
\includegraphics[width=0.95\columnwidth]{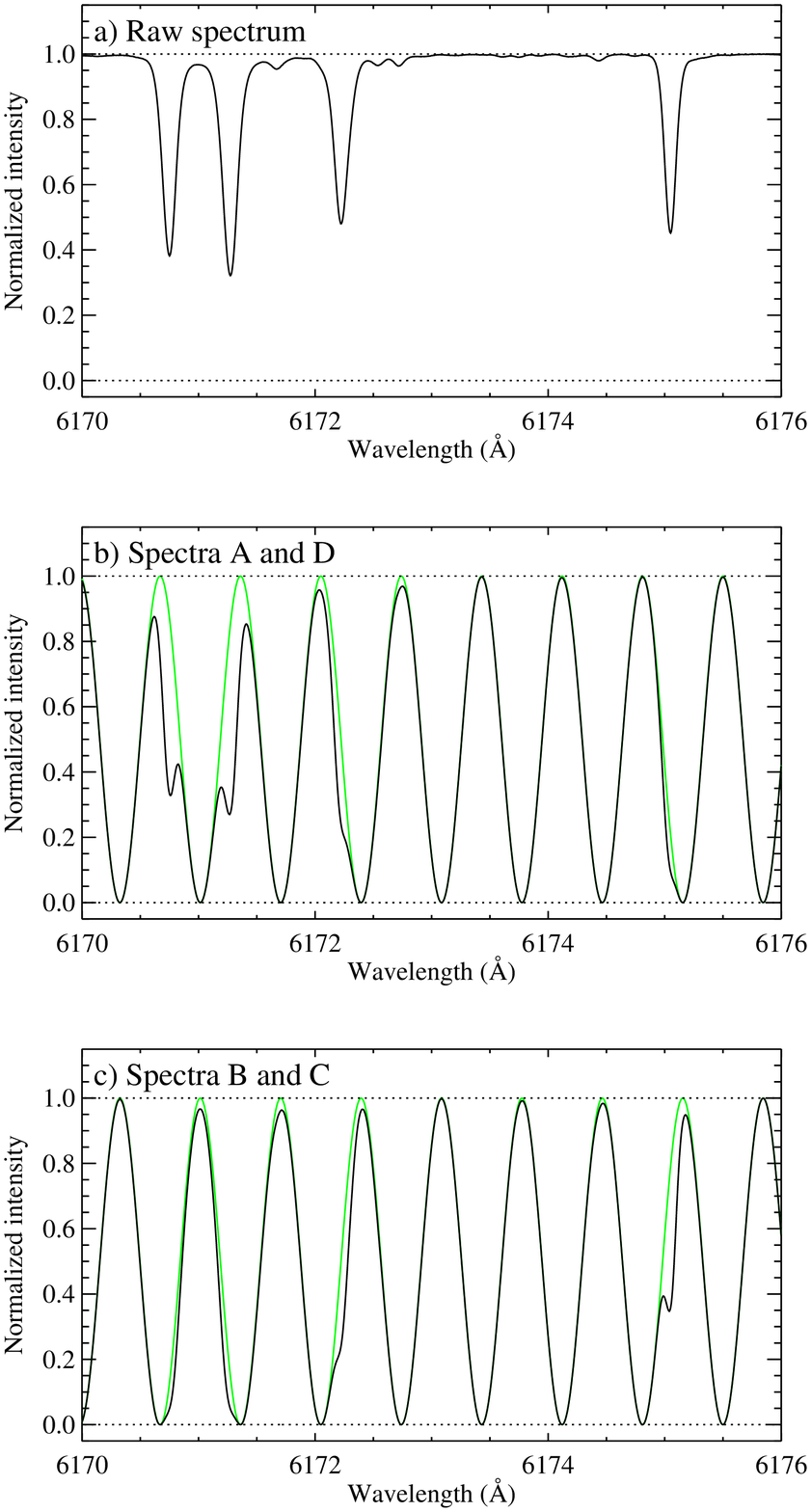}
\end{center}
\caption[]{
Top panel: A small section of a high resolution FTS spectrum of the Sun observed by \cite{2016A&A...587A..65R}.
Middle panel: 
Result of multiplying the raw spectrum with Eq. \ref{eq:trans0} (green line) with values of $\lambda_0$ 
corresponding to A and D in Fig. \ref{fig:one-element}. 
Bottom panel: Same, but with $\lambda_0$ corresponding to and B and C.
The FSR of the HMI E1 element and a contrast $B=100\%$ were used.
The FeI line around 6175~\AA\ is the one used by HMI.
}
\label{fig:spec1}
\end{figure}

With a desired Doppler velocity noise level much below 1~m/s, photon noise will dominate over readout noise.
At larger noise levels (lower signal) the readout noise will become significant, with the exact numbers
depending on observing strategy, wavelength coverage, etc.
For the present discussion I will assume that photon noise is dominant, but given a more detailed design, the accuracy of this should be revisited.
With this assumption it is straightforward to estimate the noise performance, given a high resolution spectrum and a model of the instrument.
To do so a model of the spectra is made (see Fig. \ref{fig:spec1}), including the finite spectral resolution, this model is linearized with respect to the model parameters (in particular the Doppler velocity), the noise on the ingoing spectrum is modeled as photon noise, and the noise on the parameters is determined from a least squares fits using standard statistical methods.
For the present paper a high resolution Fourier Transform Spectrometer (FTS) spectrum of the Sun observed by \cite{2016A&A...587A..65R}, with a resolution of $\approx 10^6$, is used.
The wavelength range 6000~\AA\ to 6250~\AA\ is used here for illustrative purposes, as it is mostly free of terrestrial blends and contains the spectral line used by HMI.
In all cases the performance is shown relative to an ideal spectrogram (infinite resolution) receiving the same number of photons per unit wavelength at the focal plane.

\begin{figure}
\begin{center}
\includegraphics[angle=90,width=0.9\columnwidth]{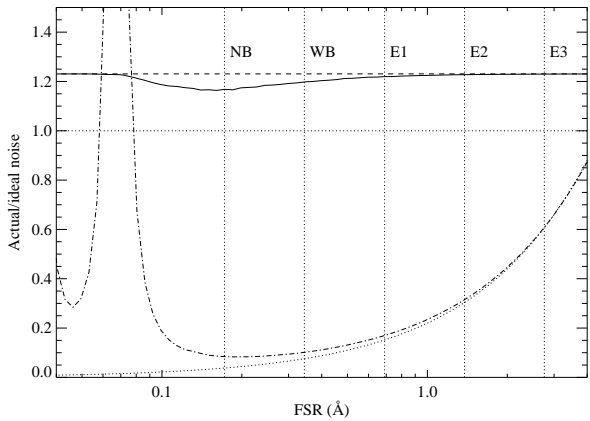}
\end{center}
\caption[]{
The noise as a function of the FSR of the inserted element for a spectrograph with a resolution of roughly R=90\,000.
Performance is shown relative to that of an ideal spectrograph with infinite resolution.
Dotted line at 1.0 is for reference.
The dashed line shows the performance of the spectrograph without the element.
The solid line the performance with the element inserted.
The dash-dotted line shows the noise on the calibration signal.
The bottom dotted line is proportional to the FSR, showing that the asymptotic behavior for high FSR is as expected.
The vertical dotted lines show the FSRs of the HMI Michelsons (NB and WB) and the narrowest three Lyot elements (E1 through E3).
For simplicity, the PSF, here and in the rest of the paper, is taken to be a boxcar, with R being the ratio of the wavelength to the width of the boxcar.
}
\label{fig:perf_both}
\end{figure}

Figure \ref{fig:perf_both} shows the performance as a function of FSR, calculated using a least squares fit, as described above.
The performance of this finite resolution spectrograph is, of course, worse than that of an ideal infinite resolution spectrograph.
At high resolution, such as shown here, the performance with a Lyot element added is only very slightly better (dip around 0.2\AA ) than an unmodified spectrograph.

\begin{figure}
\begin{center}
\includegraphics[width=0.95\columnwidth]{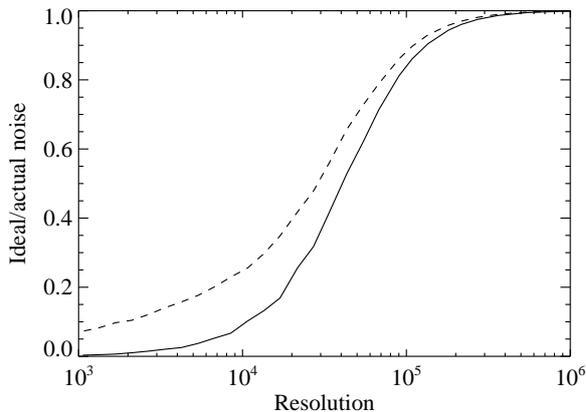}
\end{center}
\caption[]{
The velocity noise performance (defined as the ratio of the noise of an infinite resolution spectrograph to the noise of the configuration of interest) of a spectrograph without (solid)
and with (dashed) the addition of a single filter element.
For each resolution the best FSR in terms of noise
performance was selected. At high R the ideal FSR is roughly twice
the pixel size in wavelength, while at lower resolutions it is
$\approx$0.4~\AA, roughly matching the spectral line.
}
\label{fig:noise1}
\end{figure}

As all the light makes it to one of the four spectra, regardless of the contrast, the original spectrum can be recovered by summing them up. As a result the
noise performance is never worse than an unmodified spectrograph, independent of the contrast.

While the addition of the element only improves the performance slightly
at a typical resolution of R=100\,000, the improvement is substantial
at lower resolutions, as shown in 
Fig. \ref{fig:noise1}.
At a resolution of $10^4$, which is achievable with a simple
spectrograph, the improvement is about a factor of 2.5.
Indeed, this is one of the main advantages of the EDI.

\subsection{Use as a calibration signal}
\label{sec:calib}
The main use of the modulation is as a calibration signal. 
Advantages here include that the entire spectrum is modulated in
uniform way, that no light is lost (making use of all photons), that no light is added (not adding photons and thus photon noise), and
that the calibration signal is not in the form of narrow
and difficult to resolve lines.
This can be contrasted with iodine cells which do not produce
a good spectral coverage (only from about 5000~\AA\ to about 6200~\AA), have narrow lines and block some of
the light.
Similarly, ThAr lamps produce narrow lines, some very bright, and
add light (and thus photon noise) or require a separate spectrum.

An advantage of filter elements, such as the ones discussed here, is
that all wavelengths follow essentially the same path through the elements
and that the absolute position of the modulation signal is thus
described by a single number. 

To see why this is the case one needs to consider how much each of the two beam paths inside the element change with wavelength, what the temperature gradient is across this distance, and how the fringe positions change with temperature.

If we consider the setup in Fig. \ref{fig:overall}, the distance between the fiber and the element will very roughly be the same as the length of the element. As we need to offset the two images of the fiber by about one fiber width, it thus follows that the two beams will be spaced by roughly a fiber width at the output of the device or something like $ 100 \mu m$.
If the Wollaston prisms are made of quartz, the offset will vary by $\approx \pm 2 \%$ between 4000~\AA~and 8000~\AA . In other words the different wavelengths in each beam will travel within $\approx \pm 2 \mu m$ of each other or less than a part in a few thousand of the width of the collimated beam (of order 10 mm). With even a fraction of a degree difference across the device, the difference will thus be well under 0.1mK, which, as discussed in Sec. \ref{sec:temp_stab}, is adequate.
Note that temperature gradients along the rays do not enter the computation. Similarly, the difference between the two beam paths is not important, as they can be calibrated separately and averaged.

The other question is if the positions of the calibration fringes can be linearized over the relevant temperature range (some mK). To estimate this, one may use the results in Fig. 20 of \cite{Couvidat2012}. At a distance of $\Delta T$, $(d\lambda/dT)/\lambda \approx 1.4 \time 10^{-6} K^{-2} \Delta T$. In other words, the quadratic term is tiny compared to the acceptable linear drifts, discussed in Sec. \ref{sec:temp_stab}.

Given these estimates it should be a very good approximation to assume that the drift can be written as a function of wavelength times a single function of temperature.

This is unlike a regular spectrograph
where thermal expansion effects can vary across the spectrum
and ThAr lamps where the relative intensities depend on more than one factor.

The birefringent elements have the further advantage that all the
modulation occurs while the light travels through a solid material,
in other words they are insensitive to air pressure, again unlike
an in-air spectrograph or an air spaced Michelson.

Two important properties to consider are the effects of the photon noise on the calibration and that of the stability. These are discussed below.

\subsubsection{Photon noise}
\label{sec:calnoise}

The photon noise on the calibration is illustrated in Fig. \ref{fig:perf_both}.
As can be seen the calibration noise is substantially less than that
on the Doppler signal for a wide range of FSRs, including those of the HMI
Michelsons and the narrowest Lyot elements.
This is as expected, as the Doppler signal
is only sensitive to the photons in the lines, while the calibration signal uses all
wavelengths.

Unlike for the stellar signal, the noise in the calibration signal will decrease roughly in inverse proportion to a decreased contrast.
However, for HMI the contrasts are in the 96\% to 100\% range, which combined with the large margin means that this should not represent a problem.
One could potentially use this large margin to simplify the design by replacing the Wollaston prisms by partial polarizers, which would result in a single spectrum instead of four. With weak polarizers most of the light will be transmitted and the calibration signal will be weak, but strong enough for the calibration. This is more or less equivalent to inserting a weak Fabry-P\'erot and using the fringes for calibration, which may be physically simpler.

A way to overcome flatfield errors in the calibration signal is to
add a rotating 1/2~wave plate between one of the Wollaston prisms
and the birefringent element. If this is rotated by $45^\circ$
the spectra are effectively exchanged, meaning that the flat
field difference between the spectral pairs can be cancelled,
at least to lowest order.
While this might at first appear to introduce an unacceptable risk of additional
noise due to the angular repeatability of the rotation, this is
actually a small effect, as encoders can be made extremely accurate.
For HMI the rms of the repeatability of the angular position of the
rotating waveplates is
roughly $10^{-6}$ of a revolution, corresponding, for the E1 element, to an apparent
Doppler shift of
$c \times 4 \times 10^{-6} \times 0.69$~\AA$/6173$~\AA ,
or about 0.13~m/s,
where $c$ is the speed of light and
the factor 4 comes from the fact that the polarization repeats
four times when a 1/2~wave plate is rotated once.
The HMI encoders were not designed for extreme accuracy and
much better performance is available.
However, it is not clear that this added complexity is necessary and it should be pointed out that light losses and reflections from the extra waveplates need to be considered.

\subsubsection{Temperature Stability}
\label{sec:temp_stab}

As mentioned earlier, the temperature dependence of Lyot elements can be dramatically reduced by combining different materials.
At 6173~\AA\ calcite has a temperature dependence of \mbox{-17~(m/s)/mK,} KDP -80~(m/s)/mK, and ADP -350~(m/s)/mK. 
Thus the temperature dependence can be canceled by combining
a piece of calcite with a thinner piece of KDP or ADP rotated by $90^\circ$.
Generally ADP is preferable given the larger temperature dependence which means that a thinner element can be used.

The design specification for the HMI Lyot elements was 10~m\AA/K (about 0.5~(m/s)/mK).
As the temperature and wavelength dependence of the birefringence of ADP and KDP were not known accurately enough, it was decided to make test elements and measure their temperature dependence. This, unsurprisingly, resulted in thickness ratios slightly different from those determined from published values.

The actual temperature stability was measured both standalone and inside the instrument,
giving somewhat inconsistent (but within specification) numbers.
This is likely due to differences in the temperature zero points of the measurements.
Indeed, the slope appears to go to zero about 1-2 degree above the nominal operating temperature
in Fig. 20 of \cite{Couvidat2012}.
It may be noted that, as the linear term, per construction, goes to zero at the target temperature, the quadratic term dominates, especially since the measurements are done over a range of multiple K, versus the mK range relevant for the actual operation.
Using the conservative 10~m\AA/K specification and an oven stability of around 1~mK (similar to what is seen with the HMI oven) gives 0.5~m/s expected short term stability.
Regarding this number it is important to note that it is possible to control the temperature to better than 1~mK, especially as it is a small item, and that any temperature changes will be very slow.

It is important to keep in mind that the compensation is only
for bulk temperature changes (or linear temperature gradients by splitting the elements), not differential ones.
It is thus essential that temperature gradients across the elements are
minimized. For HMI this was done by surrounding the element with a conductive tube and limiting the view of other components with different temperatures.

Given that the temperature drift of ADP is much larger than that of calcite, it may be shown that in order to compensate to $\epsilon$ times the calcite drift, the thickness ratio has to be accurate to $\approx\epsilon$. In other words, the ADP thickness does not have to be extremely accurate.

\subsubsection{Wavelength dependence of sensitivity}

\begin{figure}
\begin{center}
\includegraphics[width=0.95\columnwidth]{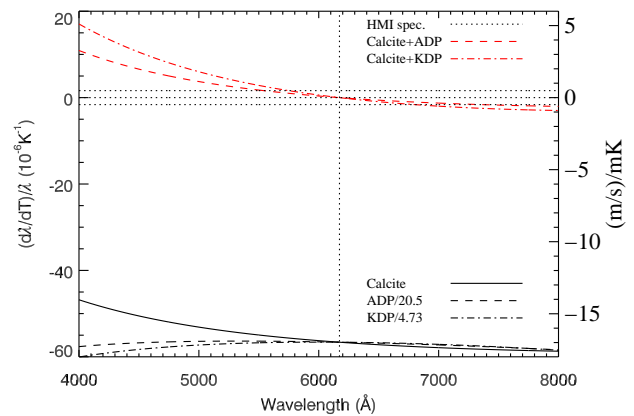}
\end{center}
\caption[]{
Wavelength dependence of the temperature dependence.
The solid black line shows the temperature dependence of calcite.
Dashed and dash-dotted black lines show ADP and KDP, scaled to having the same value as calcite at 6173~\AA. 
Top red lines show the results for calcite+ADP and calcite+KDP, optimized for 6173~\AA.
The dotted lines show the HMI specification.
The calcite numbers are from Eq. (10) of \cite{1976dbfs.rept.....T}.
The ADP and KDP numbers are derived from the 201~K and 300~K numbers in Table 1 of \cite{1982IJQE...18..143G} and 
temperature expansion coefficients of 4.2$\times 10^{-6}$ and 44$\times 10^{-6}$ for ADP and KDP, respectively \citep{Couvidat2012}.
}
\label{fig:lam_sens}
\end{figure}

The tests described above were performed at 6173~\AA\, which leaves the question of how well the temperature compensation works at other wavelengths. As already mentioned the temperature and wavelength dependence of the birefringence of ADP and KDP are poorly known.
One set of numbers were published by \cite{1982IJQE...18..143G} with the results shown in Fig.~\ref{fig:lam_sens}.
This shows a substantial 
reduction of the temperature dependence at all wavelengths. However, it is far from perfect.
Combining all three materials leads to an almost perfect cancelation, but also leads to much longer elements.
However, given the uncertainties in the published values, including those in
\cite{1982IJQE...18..143G}, in particular over which wavelength range they are reliable,
more accurate measurements of the materials are needed before a final design is made.

Depending on the results of the more detailed measurements of the various candidate materials and the desired wavelength range and stability, the required temperature stability can be determined.
If this stability is not achievable, it may be necessary to calibrate more frequently or even continuously.
 
\subsubsection{Long term stability}
\begin{figure}
\begin{center}
\includegraphics[width=0.95\columnwidth]{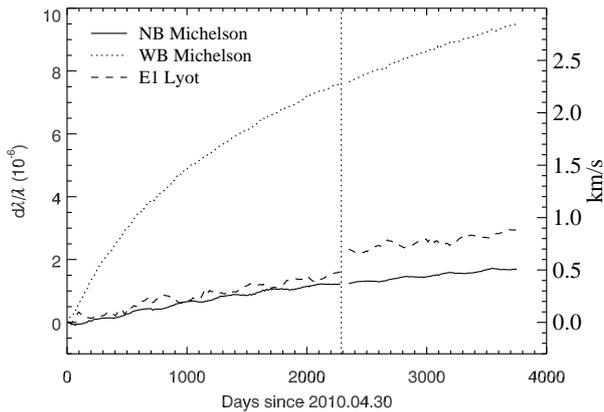}
\end{center}
\caption[]{
Change in the zero point of the three tunable elements in the HMI
instrument, as determined from the regular detunes.
Vertical line indicates the time when the instrument power was accidentally removed.
Adapted from Fig. 8 of \cite{2018SoPh..293...45H} but extended in time and with units converted. Some outliers were removed. See \cite{2018SoPh..293...45H} for details.
}
\label{fig:drift}
\end{figure}

To illustrate the long term stability of such elements, Fig. \ref{fig:drift}
shows the drift of the three tunable elements in HMI.  The drifts
are given by the peak wavelength $\lambda_0$ in Eq. (\ref{eq:trans0}) and converted to the equivalent velocity.
The physical origins of the drifts are unknown, but might be
due to such effects as radiation damage to the glass in the Michelsons
or settling of the glue used to assemble them.
For the Lyot element, in particular, the apparent drift
is likely spurious and due to the drift in other preceding filter elements, such as the front window or the dielectric blocking filter before the Lyot.
The annual variations and the jump at the power cycling of the instrument are particularly difficult to understand as the
filters are in a temperature controlled oven, again suggesting that they
do not reflect the Lyot.
Instead they may be
due to interactions of the radial velocity variations of the Sun
and filter imperfections.
Either way, the average drift of the Lyot element is 0.19~(m/s)/d before the power cycling and 0.13~(m/s)/d after,
indicating that calibration spectra need only be taken infrequently such as
once per night.

\subsubsection{Absolute calibration of the calibration signal}
\label{sec:abs}

As previously mentioned, the proposed method does not provide an absolute calibration and drifts due to residual temperature changes and long term drifts need to be corrected for.
To this end one may use the fact that the wavelength shift caused by a temperature change
is a product of a function of wavelength and a function of time, as discussed in Sec. \ref{sec:calib}.
The function of wavelength can be determined once and for all and the function of time can be determined continuously.

The simplest approach to obtaining the time dependent component is likely to add the calibration device to a spectrograph with an iodine cell, possibly modified to have weaker absorption lines than normal. The iodine spectrum then provides the absolute calibration and the calibration device extends that to the part of the spectrum not covered by the iodine lines.
A similar approach may, of course, also be used with ThAr light in separate spectra, but this negates the advantage of having a common optical path for the stellar light and the calibration signal. It also requires $2 \times 4 = 8$ sub-spectra per order.

Since the drifts are expected to be very slow it is tempting to consider only calibrating infrequently, such as at the beginning and end of each observing night. That way the light loss from, e.g. an iodine cell can be avoided, but whether the calibration can be turned on and off without disturbing the absolute calibration will need to be determined.

Another possibility would be to add the absolute calibration signal in an unused part of the spectrum, such as in the ultraviolet, infrared or parts of the spectra heavily contaminated by telluric lines.

\subsection{Disadvantages}
While the proposed design has several advantages, it also has
some disadvantages.  One obvious disadvantage is that the complexity of the instrument
is increased. The hope is that this is more than compensated by
the ability to calibrate the entire spectrum consistently and possibly reducing the requirements on the inherent stability of the instrument.
Also, it may be offset somewhat by reducing the resolution and thus
size of the spectrograph.

Four separate spectra are now generated, compared to one for the iodine approach and two for the ThAr and laser comb approaches, requiring
a larger CCD area. To some extent this is compensated by the
possible reduction in the spectrograph
resolution, but that improvement is nowhere near a factor of four or even two.
In addition to the increased CCD area, the extra spectra will also increase the impact of readout noise, which may or may not represent a significant issue.

It should be straightforward to reduce the factor to three, by overlapping
the spectra B and C in Fig. \ref{fig:one-element}.
Reducing the number of spectra to two is substantially more difficult.
One solution would be to introduce an additional amount of birefringence in one of the two beams exiting the first Wollaston prism such that they have approximately the same polarization entering the birefringent element. 
Using the angular dependence of birefringence and using one straight beam and another offset by a large angle or tilting both beams by a large angle can both achieve this.
However, to overlap the resulting pairs of spectra, these two beams must be parallel through the device and it is unclear how this can simultaneously be achieved.

In the present design the two beams are almost completely overlapping and so using the spatial separation is likely impossible. 
Separating the two beams completely should be straightforward 
using a combination of polarizing beamsplitters and mirrors. 
Unfortunately it is unclear how these two beams can be made to overlap afterward. If this is not done, the beam size will be doubled in one direction, which will have massive impacts on the downstream spectrograph.

In summary
it is not clear that a
practical arrangement can be found to reduce the number of spectra to two.
However, given the significant potential benefits, this possibility should clearly be investigated in more detail.

Imaging Michelsons can in principle be constructed in large sizes, but they are tricky to manufacture, require a different optical layout due to the bending of the light, and most likely will have to be kept in vacuum to avoid pressure effects.

Finally new calibration algorithms will have to be developed.
Experiences from using Fabry-P\'erot systems may be applicable for this, see, for example, \cite{2015A&A...581A.117B}.

\section{An extended design}
\label{sec:complicated}

A possible way to extend the design is to place a series of elements
in series, similar to what is done in a Lyot filter.
The most straightforward arrangement has factors of 2.0 between the FSRs, like in a Lyot filter.
Wollaston prisms would be placed before the first element, between them, and after the last one, with the deviation angle doubled between each.
The result would be $2^{N+1}$ spectra, with $N$ being the number of elements.
With $N=2$ elements, one would see 4 pairs of spectra (as for the $N=1$ case previously considered, there will be pairs of identical spectra) multiplied by
\begin{align}
(1+\cos(x)) & (1+\cos(2*x)) \nonumber\\
(1+\cos(x+\pi)) & (1+\cos(2*x)) \nonumber\\
(1+\cos(x)) & (1+\cos(2*x+\pi)) \nonumber\\
(1+\cos(x+\pi)) & (1+\cos(2*x+\pi)), \nonumber
\end{align}
where $x=2\pi(\lambda-\lambda_0)/{\rm FSR0}$, FSR0 is the FSR of the first element, and the contrasts have been assumed to be unity.

\begin{figure}
\begin{center}
\includegraphics[width=0.95\columnwidth]{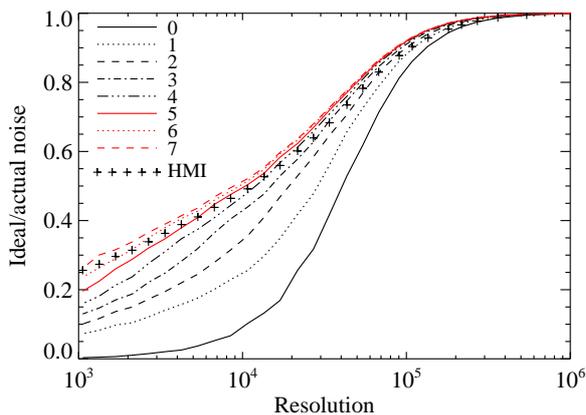}
\end{center}
\caption[]{
As Fig. \ref{fig:noise1}, but for 
various numbers of elements, as given by the legend.
The FSRs increase by a factor of two between the elements and
the FSR of the longest element is chosen to be optimal separately
for each resolution and number of elements.
Also shown with plusses is the performance of the HMI elements
with all the polarizers replaced by Wollaston prisms.
}
\label{fig:comp-perf}
\end{figure}

Fig. \ref{fig:comp-perf} shows the performance of such a system
with various numbers of elements.
As can be seen the performance indeed improves as the number of
elements is increased.

Unfortunately the performance stops increasing as more elements are
added ending with an efficiency significantly below unity.
In other words it is not possible to use this idea to make a near ideal instrument
without a large spectrograph resolution.

The required FSR to achieve the best performance decreases with both
increasing resolution and increasing number of elements.
However, with many elements the dependence of the performance on the
FSR is quite weak, as long as the FSR resolves the line.
With 7 elements the ideal FSR is roughly 0.06~\AA\ at R=$10^4$ and
0.02~\AA\ at R=$10^5$. Yet, as illustrated in Fig. \ref{fig:comp-perf},
the performance is almost as good at the FSR of HMI, which is 0.172~\AA.
This also illustrates that a buildable set of elements achieves
a good performance. Essentially the only change required to the
HMI filters would be to replace the polarizers between the elements
by Wollaston prisms.

\section{Conclusion}
\label{sec:conclusion}
Birefringent elements and Michelson interferometers represent a promising way to calibrate cross-dispersed echelle spectrographs.
Based on experience from solar instrumentation, it has been shown that the stability of such filters is adequate for calibration at the 0.1~m/s level over a substantial wavelength range.

While the approach shows great promise, an actual spectrograph has not been constructed and there are still a number of outstanding issues and undoubtedly many details that need to be worked out.
One of the first issues to address will be to obtain accurate estimates of the temperature and wavelength dependence of the birefringence for the various candidate materials.
Similarly the uniformity of the optical properties, including the birefringence, need to be measured.

Simply adding this device to an existing spectrograph is not likely to be possible, as there is not enough space between the orders to accommodate four spectra.
Having said that, it may be possible is to make a proof of concept using an existing spectrograph. If one of the Wollaston prisms is replaced by a simple polarizer, only two spectra will be generated, which can likely be accommodated in a spectrograph using a parallel calibration spectrum.

With these measurements and the proof of concept it should be possible to decide on various tradeoffs and make a more detailed design. Hopefully having eliminated any problems that have not yet been realized.

\begin{acknowledgements}
The author would like to thank
Matthias Ammler-von Eiff, Tim Brown, Patrick Gaulme, Jack Harvey, Rick Hessman, Greg Kopp, Martin Bo Nielsen, Andreas Quierrenbach, Ansgar Reiners, Dick Shine, and the anonymous referee
for useful suggestions, discussions and help with various datasets.
\end{acknowledgements}


\bibliographystyle{aa}
\bibliography{fib}

\begin{appendix}

\section{Practicalities}
\label{practical}

The use of birefringent elements and imaging Michelsons is not common. Therefore I will discuss a number of details in this Appendix, both ones relevant to the assembly of such elements and how they may be incorporated into a spectrograph.

\subsection{Placement and size}

One might think that the element can be placed almost anywhere in the optical path,
but unfortunately both types of elements have an angular dependence of $\lambda_0$.
For simple elements, like those described here, the dependence is quadratic in angle and thus focal ratio.

A calcite birefringent element will change by about $\pm$3~\AA\ across an f/10 beam, which will wash out the modulation at the FSRs considered here. As discussed in \cite{Couvidat2012} this can be improved substantially, by splitting the element in two, with a 1/2 wave plate in between, resulting in a 0.15~\AA\ shift from the center to the edge of an f/10 beam. However, as it is difficult to make waveplates perfectly achromatic, the reduction will be imperfect at some wavelengths.

For an imaging Michelson, the uncompensated shift is similar to the birefringent elements.
However, in this case the angular dependence can be reduced to fourth order by a suitable choice of refractive indices and leg lengths, thereby making the effect negligible, even for fast beams. A downside of this arrangement is that the arms must be made of different materials, leading to other problems, and the problems with making achromatic waveplates.

One solution to the angular dependence problem is to place the element in a collimated beam. Unfortunately, the beam sizes of the collimated beams in large spectrographs are generally quite large compared to the size of the elements used so far in the solar case and likely exceed the available crystal sizes. Having said that, a Michelson could potentially be made large enough.

A better solution to this problem is probably to make a short section of collimated light shortly after the fiber. This arrangement has the advantage that the beam can be made very small, which will also make the temperature control easier. 

The fast beams generally used both at the fiber output and near the camera, mean that it is almost certainly impossible to use birefringent elements there.
An imaging Michelson could potentially be made large enough and have a low enough  angular dependence to be placed elsewhere in the beam, such as near the fiber or immediately before the detector.

For a wide fielded birefringent element the angular dependence of the wavelength is
\begin{equation}
\delta v \approx \frac{c}{4 n_o^2}\frac{n_o-n_e}{n_e} \theta^2 ,
\end{equation}
where $\theta$ is the angle away from normal, $n_o$ the refractive index of the ordinary ray, and $n_e$ that of the extraordinary ray. From this it follows that the element should be mounted with an angular stability of around 0.00018 radians or 0.6 arcmin, to achieve a stability of 0.1~m/s.
For a perfectly widefielded Michelson, the leading order term is fourth order and so the required stability will be substantially reduced, likely limited by the accuracy of the widefielding.

A side effect of the proposed design is that the light falling on the Echelle will be linearly polarized. Whether this is an advantage or a disadvantage is to be determined. If circular is preferred, this can be achieved (at least approximately) by adding a 1/4 wave plate, again subject to the availability of achromatic waveplates. 
As pointed out by \cite{2015ApJ...814L..22H}, a varying polarization can lead to significant calibration problems, and so the well defined and extremely stable polarization in the present design may represent a significant advantage.

A possible problem with birefringent elements in a non-collimated beam is that they may introduce optical aberrations (astigmatism) and that the different polarizations (the four spectra) may have different imaging properties.

\subsection{Birefringent elements}

An important question to answer is what materials to use. Clearly some amount of birefringence is needed. However, it need not be very high. The HMI E1 element, which has been used for illustration here, consists of 30.62~mm calcite and 7.056~mm ADP, for a total length of less than 40~mm, which should be easy to accommodate in a spectrograph, even if mounted in a highly stable oven. Given this, other materials with somewhat lower birefringence may also be considered.

As already mentioned, the temperature and wavelength dependence of the birefringence is poorly known for many of the relevant materials. Given this, one of the first tasks will be to identify reliable measurements or to make laboratory measurements.

If large elements are needed, a serious consideration is material availability. The birefringent elements used for solar instrumentation have been about 35~mm in diameter and it is unclear if calcite crystals substantially larger than that can be obtained.
Apart from availability, the fact that calcite, ADP and KDP are soft and moderately hygroscopic, is another reason to consider alternate materials.

For temperature compensation, and if needed wide-fielding, multiple pieces need to be used. To avoid losses and reflections, it is desirable to contact these pieces. This must be done with extreme care. Not only are materials like calcite optically anisotropic, their thermal expansion is also highly anisotropic, as well as different between the materials. If two elements are glued together and subjected to a modest temperature change, the stresses can cause the crystals to break. Indeed, the original MDI Lyot filter was completely destroyed by a 10 to 15~C temperature change. To avoid this the elements can be greased (oiled) together. The redesigned MDI filter, which was assembled in this manner, was by accident subjected to more than 100~C of temperature change without breaking.

\subsection{Imaging Michelsons}

In the case of Michelsons, there are no issues with material availability for any conceivable size. However, achieving good temperature compensation does require a careful selection of materials.

A practical problem with fabricating Michelsons is the reliance on polarizing coatings and waveplates, both of which can have a significant wavelength dependence.  
At some wavelengths the coating may not split the light 50/50 into the two linear polarizations and the waveplates will deviate from 1/4 wave retardance.
One result of this is that the contrast will decrease. This is not necessarily a significant problem, as this will mostly affect the calibration noise and as there is a large margin. A more significant problem is that some of the light may be lost by being reflected back out the entrance face of the Michelson, resulting in increased noise.
The extent to which carefully designed coatings and wide-band waveplates can reduce this problem will need careful modeling.

\end{appendix}
\end{document}